\documentclass[useAMS,usenatbib]{mn2e}

\usepackage{graphicx}
\usepackage{epsfig}
\usepackage{amssymb,amsmath}
\usepackage{mathrsfs}
\usepackage{times}
\usepackage{setspace}
\usepackage{color}

\def\aj{AJ}%
\def\apj{ApJ}%
\def\apjl{ApJ}%
\def\apjs{ApJS}%
\def\aap{A\&A}%
\def\mnras{MNRAS}%
\newcommand{\begit}{\begin{itemize}}
\newcommand{\enit}{\end{itemize}}
\newcommand{\begen}{\begin{enumerate}}
\newcommand{\enen}{\end{enumerate}}

\newcommand{\beq}{\begin{equation}}
\newcommand{\eeq}{\end{equation}}
\newcommand{\beqa}{\begin{eqnarray}} 
\newcommand{\eeqa}{\end{eqnarray}} 
\def\lesssim{\mathrel{\hbox{\rlap{\hbox{\lower5pt\hbox{$\sim$}}}\hbox{$<$}}}}
\def\gtrsim{\mathrel{\hbox{\rlap{\hbox{\lower5pt\hbox{$\sim$}}}\hbox{$>$}}}}

\title[Momentum Driven Winds from Turbulence]
{Sub-Eddington Star-Forming Regions are Super-Eddington: 
Momentum Driven Outflows from Supersonic Turbulence}

\author[Thompson \& Krumholz]{
Todd A.~Thompson$^1$ \& Mark R.~Krumholz$^2$\\
$^1$Department of Astronomy and Center for Cosmology \& Astro-Particle Physics,
The Ohio State University, Columbus, Ohio 43210 \\ 
$^2$Department of Astronomy and Astrophysics, University of California, Santa Cruz, CA 95064, USA}

\begin{document}
\maketitle
\label{firstpage}
\begin{abstract}

We show that the turbulent gas in the star-forming regions of galaxies is unstable to wind formation via momentum deposition by radiation pressure or other momentum sources like supernova explosions, even if the system is below the average  Eddington limit.  This conclusion follows from the fact that the critical momentum injection rate per unit mass for unbinding gas from a self-gravitating system is proportional to the gas surface density and that a turbulent medium presents a broad distribution of column densities to the sources.   For an average Eddington ratio of $\langle\Gamma\rangle\simeq0.1$ and for turbulent Mach numbers $\gtrsim30$, we find that $\sim1$\,\% of the gas is ejected per dynamical timescale at velocities larger than the local escape velocity.  Because of the lognormal shape of the surface density distribution, the mass loss rate is highly sensitive to the average Eddington ratio, reaching $\sim20-40$\,\% of the gas mass per dynamical time for $\langle\Gamma\rangle\simeq1$.  Using this model we find a large scatter in the mass-loading factor for star-forming galaxies, ranging from $\sim10^{-3}-10$, but with significant uncertainties. Implications for the efficiency of star formation in giant molecular clouds are highlighted. For radiation pressure feedback alone, we find an increasing star formation efficiency as a function of initial gas surface density. Uncertainties are discussed.

\end{abstract}


\section{Introduction}
\label{section:introduction}

Momentum deposition in the interstellar medium (ISM) by radiation pressure on dusty gas and/or supernovae has been discussed as a mechanism for driving turbulence and launching winds in rapidly star-forming galaxies (e.g., \citealt{harwit,scoville,mqt,tqm,mmt,hopkins12_wind,zhang_thompson,ostriker_shetty,shetty_ostriker,faucher-giguere_feedback}), and in disrupting the giant molecular clouds (GMCs) around forming star clusters (e.g., \citealt{odell,scoville01,krumholz_matzner,mqt10,fall10,krumholz_dekel10,mmt,dekel13,skinner_ostriker}).  Virtually all semi-analytic treatments to date calculate the dynamics by considering the interaction between a gaseous medium with a specified set of mean properties. In these models, the behavior of the system is primarily determined by $\langle \Gamma \rangle$, the average Eddington ratio that describes the balance between momentum injection and gravity.

\cite{tqm} compared predictions for radiation pressure --- plus supernova --- supported interstellar media with observations of ultra-luminous infrared galaxies (ULIRGs) and found that the observed fluxes were close to the theoretical predictions on $\lesssim200-300$\,pc scales in these extreme systems.  \cite{andrews_thompson} compared the dusty Eddington limit with data for a broad range of galaxies, including galaxy-averaged observations of normal spirals and starbursts, individual subregions of resolved galaxies in the local universe, and ULIRGs at high redshift.  They showed that normal galaxies in general fall below $\langle\Gamma\rangle\sim1$ and that the Eddington limit presents an upper bound to the fluxes observed.  \cite{andrews_thompson} also showed that  inferences about whether or not galaxies as a whole reach the Eddington limit are hampered by uncertainties in the CO/HCN-H$_2$ conversion factors and dust-to-gas ratio (see also \citealt{faucher-giguere_feedback}), and by time-dependent effects in big spirals where the majority of the area is not star forming at a given time.    They found that $\langle\Gamma\rangle$ generally {\it decreased} for higher average gas surface density galaxies  and (similarly) that $\langle\Gamma\rangle$ increases as a function of galactocentric radius in galaxies, generically because the gas surface density falls.  This behavior follows from the fact that the ``single-scattering Eddington flux'' increases more rapidly with gas surface density than does the observed bolometric flux from galaxies.\footnote{This is equivalent to the statement that observational determinations of the star formation rate per unit area as a function of gas surface density generally find that ${\rm SFR/Area}\propto\Sigma^x$ with $x<2$ (e.g., $x\simeq1.4$; \citealt{kennicutt98}).  See Section \ref{section:galaxies}.}  Recent work by \cite{coker} on the wind of M82 also shows that it is sub-Eddington on kpc scales along its minor axis, although its super star clusters may reach or exceed $\langle\Gamma\rangle\sim1$ on small scales \citep{krumholz_matzner,mqt10,mmt}.

While this work has yielded useful insights, it has been limited to examining the mean properties of star-forming systems. However, the real ISM is far from uniform.  An important outstanding question is how momentum injection --- whether deposited by radiation pressure on dust, supernova explosions, or other processes --- couples to a supersonically turbulent ISM or GMC.  Here, we highlight an important piece of physics: over a broad range of parameters, the Eddington luminosity per unit mass ($L_{\rm Edd}/M$; in the case of radiation pressure) or the critical momentum input rate to expel gas ($\dot{P}_{\rm Edd}/M$; in the case of supernovae or stellar winds) is linearly proportional to the surface density of gas $\Sigma$ along any line of sight. For a medium with average gas surface density $\langle\Sigma\rangle$, one might imagine that if the actual luminosity $L$ or momentum injection rate $\dot{P}$ is below $\langle L_{\rm Edd}\rangle$ or $\langle\dot{P}_{\rm Edd}\rangle$ that no material is ejected. However, since the medium is supersonically turbulent, the momentum sources ``see" a broad lognormal probability distribution function (PDF) of  surface densities whose width is controlled by the Mach number of the turbulence.  At sufficiently low $\Sigma$, the  medium will thus be super-Eddington.  Setting the $\Sigma$-dependent Eddington ratio to unity defines a critical surface density $\Sigma_{\rm crit}$.  The material in sightlines with $\Sigma<\Sigma_{\rm crit}$ will be accelerated to the local escape velocity (or above) in a local dynamical timescale (or less).

We develop this model here. We argue that the gas along sightlines with $\Sigma<\Sigma_{\rm crit}$ is ejected.   The mechanism is generic in the sense that any turbulent medium with momentum sources should be unstable to some mass loss.   In practice, we show that because of the shape of the surface density and mass PDFs the system needs to be within about 1/10th of the Eddington limit for significant gas expulsion, with the degree  of mass loading  determined by the Mach number of the turbulence and the ratio $\Sigma_{\rm crit}/\langle\Sigma\rangle$.

In the context of radiation pressure on dust, a handful of studies have previously considered the interaction between radiation forces and a turbulent medium, and these have focused on the highly optically-thick limit applicable to surface densities  larger than $\sim10^4$\,M$_\odot$ pc$^{-2}$\,$\sim0.5$\,g cm$^{-2}$\,$\sim10^{23}$\,cm$^{-2}$, where the re-radiated FIR emission might be trapped by the dusty gas.  These works include both semi-analytic calculations  with model turbulent column density PDFs \citep{mqt10, hopkins11_feedback} and full numerical radiation hydrodynamics  \citep{krumholz_thompson12,krumholz_thompson13,davis_kt,skinner_ostriker} to assess whether the effective momentum coupling optical depth is in fact as large as the average FIR optical depth in a turbulent medium with lower column density sightlines.  The general result of these studies is that the actual momentum deposition is smaller than what one would guess for a laminar, non-turbulent medium, though the extent of the deviation is still uncertain. We note that in the FIR optically-thick limit, \cite{davis_kt} (their Section 5.2) discussed the possibility that radiation pressure might launch outflows from galaxies that are below the average Eddington limit by having enhanced fluxes in lower-column density channels driven by the radiation Rayleigh-Taylor instability.  Similar numerical calculations applicable to the broad column density regime where $L_{\rm Edd}/M\propto\Sigma$, and where the effect highlighted in this paper should apply have yet to be undertaken in the galaxy feedback context.  

In the supernova context, \cite{shetty_ostriker} and \cite{martizzi} have explored how the momentum deposition couples to and regulates the turbulent ISM, but neither global outflows nor GMC disruption were the focus of their work.  \cite{hopkins12_wind} explored the development of winds in full galaxy simulations with both radiation pressure and supernova feedback, and in principle, if turbulence was well-resolved on the scales of individual GMCs, we would expect the effect identified here to be present in their calculations.  However, in their simulations it is difficult to disentangle the effects of supernovae, which were treated with full hydrodynamics, from the effects of radiation pressure, which were handled via sub-grid model analogous to the semi-analytic treatments discussed above.  \cite{creasey} follow the development of outflows in a supernova-driven ISM with energy deposition and momentum.  A structured turbulent ISM was also part of \cite{cooper}'s calculation of the superwind from M82.  As we discuss in Section \ref{section:momentum}, the ram pressure acceleration of low-column density sightlines ($\Sigma<\Sigma_{\rm crit}$) by either a hot wind produced by energy injection from supernovae or the direct momentum injection can also be interpreted in terms of an Eddington limit.   The former point was recently explored by \cite{zhang15}.

In Section \ref{section:edd} we discuss the Eddington limit for momentum input in a medium of given surface density.  In Section \ref{section:turbulence} we compute the PDF of area and mass for a vertically-averaged supersonically turbulent medium and calculate the interaction of momentum sources with this medium.  In Section \ref{section:discussion} we provide a discussion of our results, with applications to the ISM of galaxies and GMCs, and other momentum sources.  Section \ref{section:conclusion} provides a brief conclusion.

\section{The Eddington Limit in Driven Supersonic Turbulence}
\label{section:edd}

\subsection{The Single-Scattering Eddington Limit}
\label{section:eddington}

We start by considering a spherical source of UV/optical luminosity $L$ and mass $M$ surrounded
by a turbulent dusty medium.  Our arguments can be generalized to a thin disk geometry or 
other momentum sources.
The medium surrounding the source has an area-averaged gas surface 
density $\langle\Sigma\rangle$, but presents a distribution of surface densities to the central
source because it is turbulent.  The equation of motion for gas along any line of sight 
with surface density $\Sigma$ that is optically-thick to the incoming emission but optically-thin to 
the reradiated FIR is
\beq
\frac{dv}{dt}=-\frac{GM}{r^2}+\frac{L}{4\pi r^2 c}\frac{1}{\Sigma},
\label{eom}
\eeq
which implies an Eddington luminosity of 
\beq
\frac{L_{\rm Edd}}{M}\simeq 4\pi G c\Sigma \simeq  130\frac{\rm L_\odot}{\rm M_\odot}\,\Sigma_{0.01},
\label{lm}
\eeq
where $\Sigma_{0.01}=\Sigma/0.01$\,g cm$^{-2}$ and
0.01\,g cm$^{-1}\simeq50$\,M$_\odot$ pc$^{-2}\simeq6\times10^{21}$\,cm$^{-2}$.  
The Eddington ratio along the line of sight is then
\beq
\Gamma(\Sigma)=\frac{L}{L_{\rm Edd}(\Sigma)},
\eeq
where we explicitly note the functional dependence of $\Gamma$ on $\Sigma$.  In the case of an arbitrary momentum source $\dot{P}$, one can substitute $L/c\rightarrow\dot{P}$ in the above.  Recent work by \cite{faucher-giguere_feedback} implies that the net momentum injection rate from supernovae might be as much as $\dot{P}_{\rm SNe}\sim10 L/c$.  We return to this issue in Section \ref{section:discussion}.

\begin{figure*}
\centerline{\includegraphics[width=1\textwidth]{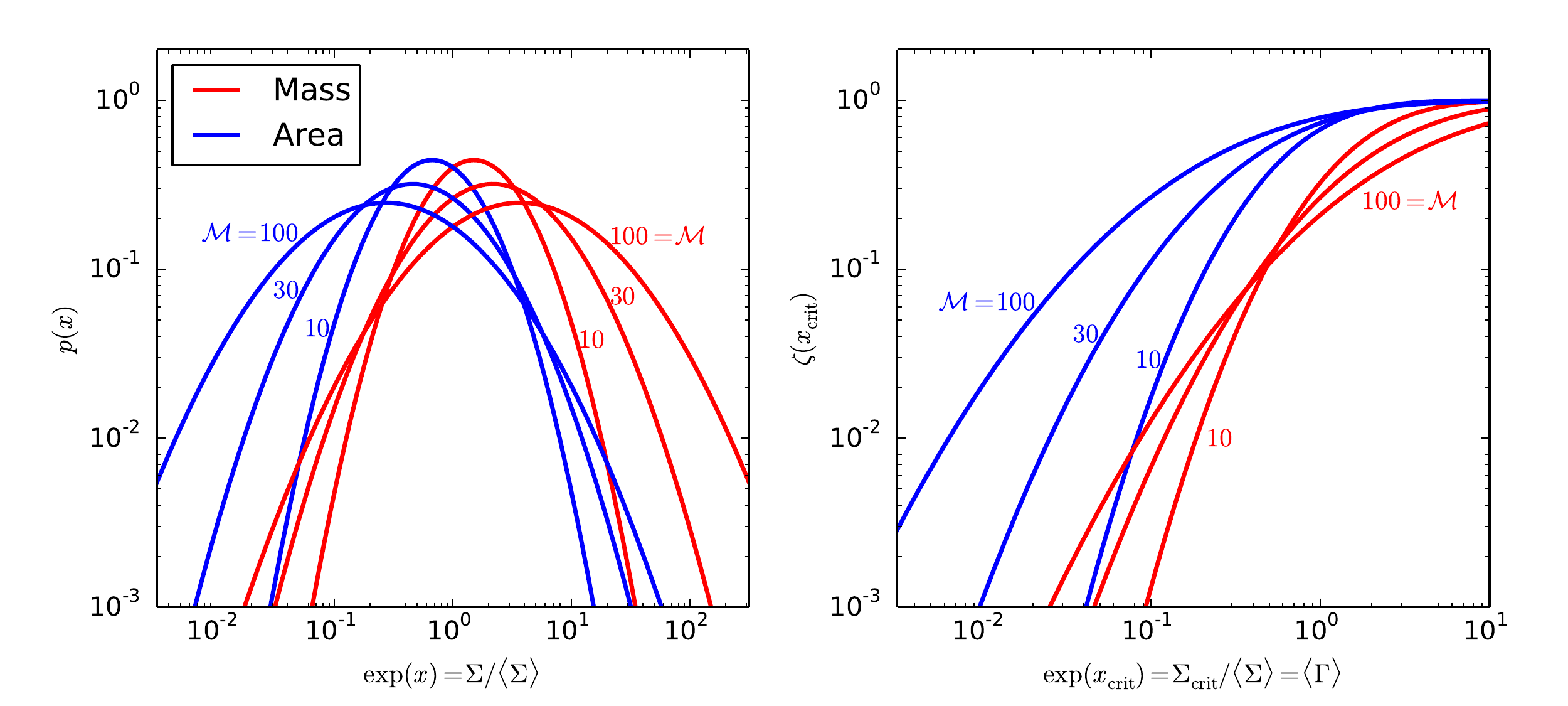}}
\caption{{\it Left:} The areal column density probability distribution function (PDF) (blue; $p_+(x)$) and the mass column density PDF (red; $p_-(x)$) as given in equation~\ref{p}, as a function of $e^x=\Sigma/\langle\Sigma\rangle$ for Mach numbers ${\cal M}=10$, 30, and 100.  This plot shows that the areal (mass) PDF is peaked below (above) the mean value of the column density, and that it broadens as ${\cal M}$ increases. {\it Right:} The integral of the areal (blue; $\zeta_+$) and mass (red; $\zeta_-$) PDFs from $x=-\infty$ to $x_{\rm crit}$ (eq.~\ref{xcrit}), as given in eq.~\ref{zeta}, as a function of the average Eddington ratio $\exp(x_{\rm crit})=\langle\Gamma\rangle$.  This panel shows that for a system with $\langle\Gamma\rangle=0.1$ and ${\cal M}\gtrsim30$ that $\gtrsim10$\% ($\zeta_+\gtrsim0.1$) of the area and $\gtrsim1$\,\% ($\zeta_-\gtrsim10^{-2}$) of the mass is super-Eddington. }
\label{figure:pdf}
\end{figure*}

In the context of radiation pressure, note that equation (\ref{eom}) is valid over about 2.5 dex in $\Sigma$.  For $\Sigma\lesssim1/\kappa_{\rm UV}\simeq10^{-3} \kappa^{-1}_{\rm UV,\,3}f_{\rm dg,\,MW}^{-1}$\,g cm$^{-2}$ --- where  $\kappa_{\rm UV,\,3}=\kappa_{\rm UV}/10^3$\,cm$^2$ g$^{-1}$ is a typical UV continuum dust opacity and $f_{\rm dg,\,MW}$ is the dust-to-gas ratio scaled to the Milky Way value --- the medium becomes optically-thin to the incident UV radiation and $1/\Sigma$ should be replaced by $\kappa_{\rm UV}$ in equation (\ref{eom}).   Conversely, for $\Sigma\gtrsim1/\kappa_{\rm IR}\simeq0.2 \kappa^{-1}_{\rm IR,\,0.7}f_{\rm dg,\,MW}^{-1}$\,g cm$^{-2}$, where $\kappa_{\rm IR}$ is the Rosseland-mean opacity (e.g., \citealt{semenov}), the medium becomes optically thick to the reradiated IR and $1/\Sigma$ should be replaced by $\kappa_{\rm IR}$ in  equation (\ref{eom}).  However, as discussed by many authors (see, e.g., \citealt{krumholz_matzner,mqt10,krumholz_thompson12,krumholz_thompson13,davis_kt}), it is unclear if the  momentum coupling in the optically-thick limit translates into a momentum input as large as $\tau_{\rm IR}=\kappa_{\rm IR}\Sigma$ in a turbulent medium.  Here, we focus on the range of parameters where the single-scattering limit applies and return to a discussion of the IR optically-thick and UV optically-thin regimes in Section \ref{section:discussion}.

If the turbulent dusty medium has an average gas surface density $\langle\Sigma\rangle$, the average Eddington luminosity is simply 
\beq
\langle L_{\rm Edd}\rangle=4\pi GMc\langle\Sigma\rangle.
\eeq
From equation (\ref{lm}) we then see that for source luminosity $L$, there is then a critical surface density below which $\Gamma\ge1$:
\beq
\frac{\Sigma_{\rm crit}}{\langle\Sigma\rangle}=\frac{L}{\langle L_{\rm Edd}\rangle}\equiv\langle\Gamma\rangle.
\label{sigmacrit}
\eeq

All regions exposed to the central source with $\Sigma<\Sigma_{\rm crit}$ will be accelerated out of the local gravitational potential by momentum deposition.
Solving equation (\ref{eom}) we find the classic result that 
\beq
\frac{v(r)}{v_{\rm esc}(R_0)}=\left(1-\frac{R_0}{r}\right)[\Gamma(\Sigma)-1]^{1/2},
\label{vinfty}
\eeq
where $v_{\rm esc}=(2GM/R_0)^{1/2}$, $R_0$ is the initial radius of the medium, and  where we have assumed that $\Sigma$ is constant with radius as the cloud is accelerated.  In this limit, for $r\gg R_0$, $v_\infty\simeq v_{\rm esc}(R_0)(\Gamma-1)^{1/2}$ and the asymptotic velocity is tied to the escape velocity from the central source. If the ejected medium instead expands as it is accelerated so that the angle subtended by the cloud from the source does not decrease as $r^{-2}$,  then the cloud may reach much higher velocity \citep{thompson_shell}.  In the  context of radiation pressure on dust, if the cloud subtends a constant solid angle as it is accelerated, its asymptotic velocity is $v_\infty\simeq(R_{\rm UV}L/(M_{\rm cloud} c))^{1/2}$ if $R_{\rm UV}\gg R_0$, where $M_{\rm cloud}$ is the mass of the cloud and $R_{\rm UV}=(\kappa_{\rm UV}M_{\rm cloud}/4\pi)^{1/2}$ is the radius at which the cloud becomes optically thin to the incident UV radiation.  This assumes that the source $L$ is constant on the timescale needed to reach $R_{\rm UV}$, which is unlikely in the case of an intervening turbulent medium.  We return to this issue in Section \ref{section:turbulence}.

Taking equation (\ref{vinfty}) at face value, the acceleration time on a scale $R_0$ is then 
\beq
t_{\rm acc}(\Sigma)=\left(\frac{R_0^3}{2GM}\right)^{1/2}\left[\Gamma(\Sigma)-1\right]^{-1/2}.
\label{tacc}
\eeq

\subsection{Interaction with a Turbulent Medium}
\label{section:turbulence}

In a turbulent medium, the sources of luminosity see a broad distribution of 
column densities along each line of sight, and the fraction of mass that finds itself super-Eddington will depend on this distribution. A number of numerical experiments have found that, for supersonic isothermal turbulence, the probability distribution function (PDF) of column densities is well-approximated by a lognormal distribution \citep{ostriker01, vazquez01, federrath10}
\beq
p_\pm (x)=\frac{1}{(2\pi\sigma_{\ln \Sigma}^2)^{1/2}}\exp\left[-\frac{(x-\overline{x})^2}{2\sigma_{\ln \Sigma}^2}\right]
\label{p}
\eeq
where $x = \ln (\Sigma/\langle\Sigma\rangle)$. Conservation of mass requires that the mean $\overline{x}$ and dispersion $\sigma_{\ln\Sigma}$ be related by $\overline{x}=\mp\sigma_{\ln\Sigma}^2/2$. The quantity $p_+(x)$ gives the areal PDF (i.e., the probability of measuring a certain column density if one chooses a line of sight passing through a random position), while $p_-(x)$ gives the mass PDF (i.e., the probability of measuring a certain column density if one chooses a line of sight passing through a random mass element).

The dispersion of the lognormal $\sigma_{\ln \Sigma}$ is related to the Mach number of the turbulence. For volume density, which is also well-described by a lognormal, numerous authors have found that the non-magnetized turbulence with mixed forcing produces a dispersion
\beq
\sigma_{\ln \rho}^2 \approx \ln (1+\mathcal{M}^2/4)
\eeq
in the log density PDF, where $\mathcal{M}$ is the three-dimensional Mach number of the turbulence \citep[e.g., see the recent review by][]{krumholz14_review}. The dispersion of the column density PDF is smaller due to averaging over the line of sight, but the relationship has been subject to significantly less exploration in numerical simulation than that between $\mathcal{M}$ and the dispersion of the volume density PDF. To estimate the dispersion of the column density PDF, we follow an approach suggested by \citet{brunt10a, brunt10b}. They consider a periodic box of size $L$, and show that the ratio of the dispersions of column density, $\sigma_\Sigma$, and volume density, $\sigma_\rho$, are related by
\beq
\label{eq:bfp_rfactor}
R \equiv \left(\frac{\sigma_\Sigma}{\sigma_\rho}\right)^2 =
\frac{\sum_{k_x, k_y=-\infty}^{\infty} P(k_x,k_y,0) - P(0,0,0)}
{\sum_{k_x, k_y, k_z=-\infty}^{\infty} P(k_x,k_y,k_z) - P(0,0,0)},
\eeq
where $P(k_x,k_y,k_z)$ is the power spectral density of the density field at point $(k_x,k_y,k_z)$ in Fourier space, and the wave vectors $(k_x,k_y,k_z)$ are normalized such that $k_x = 1$ corresponds to a mode with wavelength $\lambda = 2L$, i.e., $k_x = 1$ is the largest mode that will fit in the box. Note that these are dispersions of the column and volume densities themselves, not of their logarithms. For a lognormal PDF, the dispersions of the density and its logarithm are related by
\begin{equation}
\sigma_{\ln \rho}^2 = \ln (1+\sigma_\rho^2),
\end{equation}
and similarly for the column density. Combining the previous three equations, we have
\beq
\label{eq:coldispersion}
\sigma_{\ln \Sigma}^2 \approx \ln (1 + R \mathcal{M}^2/4).
\eeq

The value of $R$ depends on the shape of the density power spectrum. For homogenous isotropic turbulence $P(k_x,k_y,k_z)$ must depend on $k = (k_x^2+k_y^2+k_z^2)^{1/2}$ alone, and several numerical studies have found that the relationship is reasonably well-described by a power law $P(k) \propto k^{-\alpha}$ with an index $\alpha$ that depends on the Mach number of the turbulence, and varies form $\alpha \approx 3.7$ at near-transonic turbulence to $\alpha \approx 2.5$ for $\mathcal{M} \gg 1$ (\citealt{kim05, beresnyak05}; for a summary and references to further results, see the review by \citealt{krumholz14_review}). Since most of the astrophysical systems with which we are concerned have $\mathcal{M} \gg 1$, we adopt $\alpha=2.5$ as our fiducial value. This power law behavior of $P(k)$ must stop at sufficiently high $k$ for the dispersion to remain finite, and the natural truncation scale is the sonic length scale, below which the turbulence becomes subsonic and is thus no longer able to drive density fluctuations \citep{krumholz05}; in our normalized wave vector units, this scale corresponds to $k = \mathcal{M}^2$. We therefore adopt as an \textit{ansatz} that 
\beq
\label{eq:pofk}
P(k) \propto 
\left\{
\begin{array}{ll}
0, & k = 0 \\
k^{-\alpha}, & 1 \leq k \leq \mathcal{M}^2 \\
0, & k > \mathcal{M}^2
\end{array}
\right..
\eeq
With this \textit{ansatz}, and approximating the sums in equation (\ref{eq:bfp_rfactor}) by integrals (appropriate for $\mathcal{M} \gg 1$), we have
\beq
R = \frac{1}{2}\left(\frac{3-\alpha}{2-\alpha}\right)\left[\frac{1 - \mathcal{M}^{2(2-\alpha)}}{1 - \mathcal{M}^{2(3-\alpha)}}\right].
\eeq
By using this value of $R$ in equation (\ref{eq:coldispersion}) we have completed the specification of the PDF of $\Sigma$ in terms of $\mathcal{M}$.

Now we are in a position to ask how much mass and area is contained in regions where $\Sigma$ is small enough for the gas to be super-Eddington. We define a critical value of $x$ for this condition to be satisfied as (see eq.~\ref{sigmacrit})
\beq
x_{\rm crit}=\ln[\Sigma_{\rm crit}/\langle\Sigma\rangle].
\label{xcrit}
\eeq
Integrating $p_\pm(x)$ from
$x=-\infty$ to $x=x_{\rm crit}$ yields the total fraction of the area and mass, respectively,  
of the medium with $\Sigma<\Sigma_{g,\,\rm crit}$:
\beqa
\zeta_\pm(x_{\rm crit})&=&\int_{-\infty}^{x_{\rm crit}} p_\pm(x) \,dx \nonumber \\
&=&\frac{1}{2}\left[1\pm{\rm erf}\left(\frac{\pm 2x_{\rm crit}+\sigma_{\ln\Sigma}^2}{2\sqrt{2}\sigma_{\ln\Sigma}}\right)\right].
\label{zeta}
\eeqa
In the right hand panel of Figure \ref{figure:pdf} we plot $\zeta_\pm(x_{\rm crit})$ for the area (blue) and the mass (red), respectively, for ${\cal M}=10$, 30, and 100.  For $\langle\Gamma\rangle\simeq0.1$  and  ${\cal M}=30$, we see that $\zeta_-\sim10^{-2}$ and $\zeta_+\sim0.2$, implying that about $1$\% of the mass and $20$\% of the area of the system would be super-Eddington.

An important question is whether the matter along a super-Eddington line of sight can be accelerated before the turbulence ``erases" the local conditions.  If the column density field fluctuates in a time much less than the time to accelerate the matter, we expect no material to be ejected.  The relevant comparison is then the ratio $t_{\rm acc}(\Sigma)/t_{\rm cross}(\lambda)$, where $t_{\rm acc}$ is given by equation (\ref{tacc}) and $t_{\rm cross}(\lambda)\sim \lambda/\delta v(\lambda)$ is the crossing time of the turbulence with velocity $\delta v(\lambda)$ to cross a scale $\lambda$ over which $\Sigma$ obtains: 
\beq
\frac{t_{\rm acc}(\Sigma)}{t_{\rm cross}(\lambda)}\sim
\left(\frac{\delta v(\lambda)}{v_{\rm esc}(R_0)}\right)\left(\frac{R_0}{\lambda}\right)
\frac{1}{[\Gamma(\Sigma)-1]^{1/2}},
\eeq
where $v_{\rm esc}(R_0)=(2GM/R_0)^{1/2}$.  To our knowledge, the projected persistence time of low column density structures in simulations of highly supersonic turbulence has not been reported in the literature.  However, in the GMC context on the largest scales ($\lambda\sim R_0$), we expect $\delta v\sim v_{\rm esc}$, implying that $t_{\rm acc}(\Sigma)<t_{\rm cross}(R_0)$ if $\Gamma(\Sigma)\gtrsim{\rm few}$.  A similar conclusion is reached by considering a geometrically thin disk with flux $F$ and total surface density $\Sigma_{\rm tot}$ on a scale height $h$. We thus conclude that for $\Gamma(\Sigma)$ larger than $\sim2$, the material should be accelerated before turbulence erases the local conditions.  In the GMC context, the material is accelerated to the escape velocity, whereas for a geometrically thin disk the super-Eddington matter will be accelerated to a characteristic velocity of $v^2\sim \pi G\Sigma_{\rm tot} h$, the vertical velocity dispersion of the gas.\footnote{This follows from solving the one-dimensional planar momentum equation for a column of gas being accelerated out of a thin disk (see eq.~\ref{eom}), or by equating $\rho v^2/2\sim \pi G\Sigma_{\rm tot}^2$ in hydrostatic equilibrium and noting $\Sigma_{\rm tot}=2\rho h$.}

\begin{figure}
\centerline{\includegraphics[width=8.5cm]{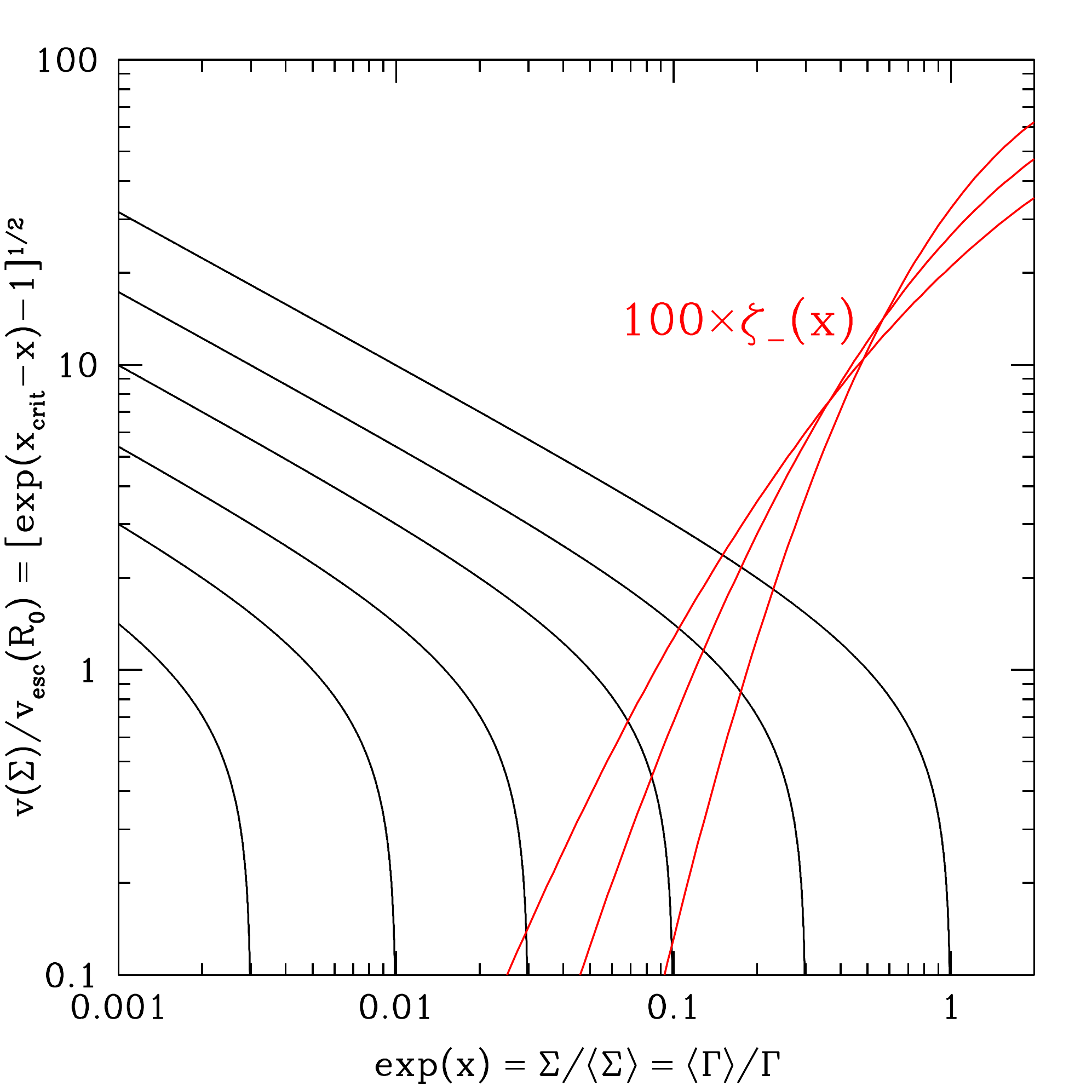}}
\caption{Velocity distribution $v(\Sigma)/v_{\rm esc}(R_0)$ (black lines) of ejected material as a function of $e^x=\Sigma/\langle\Sigma\rangle=\langle\Gamma\rangle/\Gamma$ for $\langle\Gamma\rangle=3\times10^{-2}$, 0.01, 0.03, 0.1, 0.3, and 1 (lowest to highest).  The red lines show $100\zeta_-(x)$ from Figure \ref{figure:pdf} (right panel) for ${\cal M}=10$, 30, and 100. See equation (\ref{vdist}).}
\label{figure:v}
\end{figure}

If the material with $\Sigma<\Sigma_{\rm crit}$ is ejected, and if the surface density of the ejected regions is constant as a function of radius in its first dynamical time as it  is accelerated, then the velocity distribution is just 
\beq
\frac{v(\Sigma)}{v_{\rm esc}(R_0)}\simeq\left(\frac{\Sigma_{\rm crit}}{\Sigma}-1\right)^{1/2}
=\left[e^{(x_{\rm crit}-x)}-1\right]^{1/2}.
\label{vdist}
\eeq
Figure \ref{figure:v} shows $v/v_{\rm esc}$ as a function of $e^x$ for a wide range of $\langle\Gamma\rangle$ from $3\times10^{-2}-1$ (black solid lines, lowest to highest).  The red lines show the integral of the mass PDF times 100 ($100\zeta_-(x)$) from the right panel of Figure \ref{figure:pdf}.  We see that only a very small amount of mass can reach $v/v_{\rm esc}\gg1$ in a single dynamical time on scale $R_0$.   For example, taking $\langle\Gamma\rangle=0.1$, a fraction $\sim10^{-2}$ ($100\zeta_-\simeq1$) of the mass reaches $v/v_{\rm esc}\sim1$ for ${\cal M}=100$, but only $\lesssim10^{-3}$ reaches $v/v_{\rm esc}\gtrsim2$ because of the very strong drop in $\zeta_-(x)$.  Taking $\langle\Gamma\rangle=1$, more than 0.02 of the gas mass reaches $v/v_{\rm esc}\gtrsim2$, whereas a fraction 0.001 reaches $v/v_{\rm esc}\gtrsim6$.

\section{Discussion}
\label{section:discussion}

In Section \ref{section:galaxies} we first discuss the application of our results to observed star-forming galaxies as a whole in the context of radiation pressure. For nominal CO-H$_2$ conversion factors these systems are on average below the single-scattering Eddington limit \citep{andrews_thompson}.  Because in a thin disk geometry we expect low-column sightlines that are near the Eddington limit to be accelerated to only of order the vertical velocity dispersion of the gas ($\sim10-50$\,km s$^{-1}$ for most systems) the importance of radiation pressure alone as a momentum injection mechanism is likely not dramatic for most systems.  Even so, some galaxies do have average Eddington ratios closer to unity than others, and in all galaxies some gas will be in low-column density sightlines that are effectively strongly super-Eddington. In addition, we find that our results depend sensitively on both the CO-H$_2$ conversion factor and the total momentum injection rate, which may be contributed to by other momentum sources (see Section \ref{section:momentum}).  

The application to massive star-forming sub-regions within galaxies is given in Section \ref{section:gmcs}.  Even $z\sim0$ galaxies may reach the single-scattering radiation pressure Eddington limit \citep{krumholz_matzner,mqt10,mmt}, driving shells vertically out of the plane of galaxies at relatively high velocity (Fig.~\ref{figure:v}), and our results indicate the star-formation efficiency in GMCs may be substantially modified by the interaction of radiation pressure with the turbulent medium.  

\subsection{Application to Galaxies}
\label{section:galaxies}

In a galactic disk with a continuous star formation rate per unit area $\dot{\Sigma}_*$, the flux produced by the newborn stars will be $F \approx \epsilon \dot{\Sigma}_\star c^2$, where $\epsilon\simeq7\times10^{-4}$ is an IMF-dependent constant. This can be compared to the Eddington flux, which is
\beq
F_{\rm Edd} = 2\pi G c f_g^{-1} \Sigma_g^2,
\eeq
where $\Sigma_g$ is the gas surface density and $f_g$ is the gas mass fraction. We therefore have
\beq
\label{eq:xcritgal}
e^{x_{\rm crit}} = \frac{F}{F_{\rm Edd}} = \left(\frac{\epsilon c f_g}{2\pi G}\right) \frac{\dot{\Sigma}_*}{\Sigma_g^2}.
\eeq
Loci of constant $F/F_{\rm Edd}$ therefore correspond to relationships $\dot{\Sigma}_* \propto \Sigma_g^2$ between galaxies' star formation rate and gas content (at fixed $f_g$; see, e.g., \citealt{andrews_thompson}). Conversely, for any observed galaxy for which $\Sigma_g$ and $\dot{\Sigma}_*$ are known, and for which $f_g$ is known or measured, we can use equation (\ref{eq:xcritgal}) to infer $x_{\rm crit}$ and thence $\zeta$, the fraction of mass that is super-Eddington.

\begin{figure}
\centerline{
\includegraphics[width=0.55\textwidth]{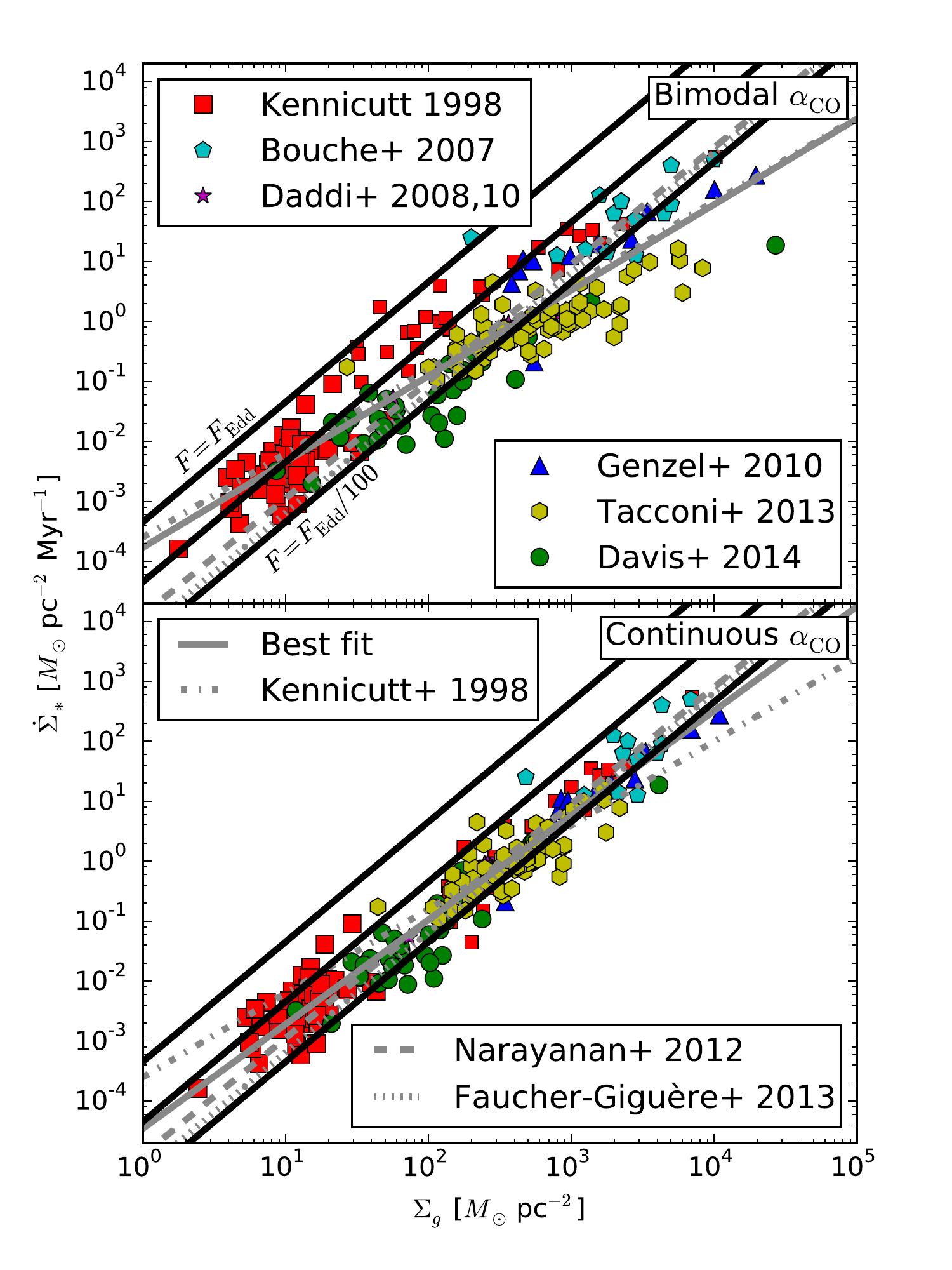}
}
\caption{
\label{fig:galaxies1}
Locations of observed galaxies, together with theoretical curves, in the $\Sigma_g - \dot{\Sigma}_*$ plane. Black lines show the locations for $F=F_{\rm Edd}$, $F = F_{\rm Edd}/10$, and $F = F_{\rm Edd}/100$, as indicated; these curves were computed for $f_g = 0.3$, and their vertical position varies as $f_g^{-1}$. Points indicate a compilation of data on observed galaxies taken from \citet{krumholz14_review}. Original sources for the data are as follows: \citet{kennicutt98}, \citet{bouche07a}, \citet{daddi08a} and \citet{daddi10a}, \citet{genzel10a}, \citet{tacconi13a}, and \citet{davis14a}. Gray curves indicate a least squares fit to this data set (solid), the \citet{kennicutt98} relation (dot-dashed), and fits by \citet{narayanan12a} (dashed) and \citet{faucher-giguere_feedback} (dot-dashed) to subsets of the data shown, and using a variety of models for $\alpha_{\rm CO}$ (see also \citealt{ostriker_shetty}). Finally, the upper panel shows the results using the bimodal $\alpha_{\rm CO}$ values recommended by \citet{daddi10b}, while the bottom panel uses the theoretical $\alpha_{\rm CO}$ value computed by \citet{narayanan12a}.
}
\end{figure}

In Figure \ref{fig:galaxies1} we show loci of constant $F/F_{\rm Edd}$ assuming a constant gas fraction of $f_g=0.3$ in the $\Sigma_g - \dot{\Sigma}_*$ plane overlaid with data for a large collection of local and high-redshift star-forming galaxies taken from the compilation of \citet{krumholz14_review}. For the assumed $f_g$, observed galaxies fall below $F = F_{\rm Edd}$, and most fall below $F = F_{\rm Edd}/10$. There is a systematic trend whereby higher star formation rate galaxies have lower values of $F/F_{\rm Edd}$, which is simply a consequence of the fact that constant $F/F_{\rm Edd}$ would require that the star formation rate rise with gas surface density as $\Sigma_g \propto \Sigma_*^2$ (e.g., \citealt{tqm,andrews_thompson}), and the observed relationship between gas content and star formation is apparently not quite that steep. 

\begin{figure}
\centerline{
\includegraphics[width=0.55\textwidth]{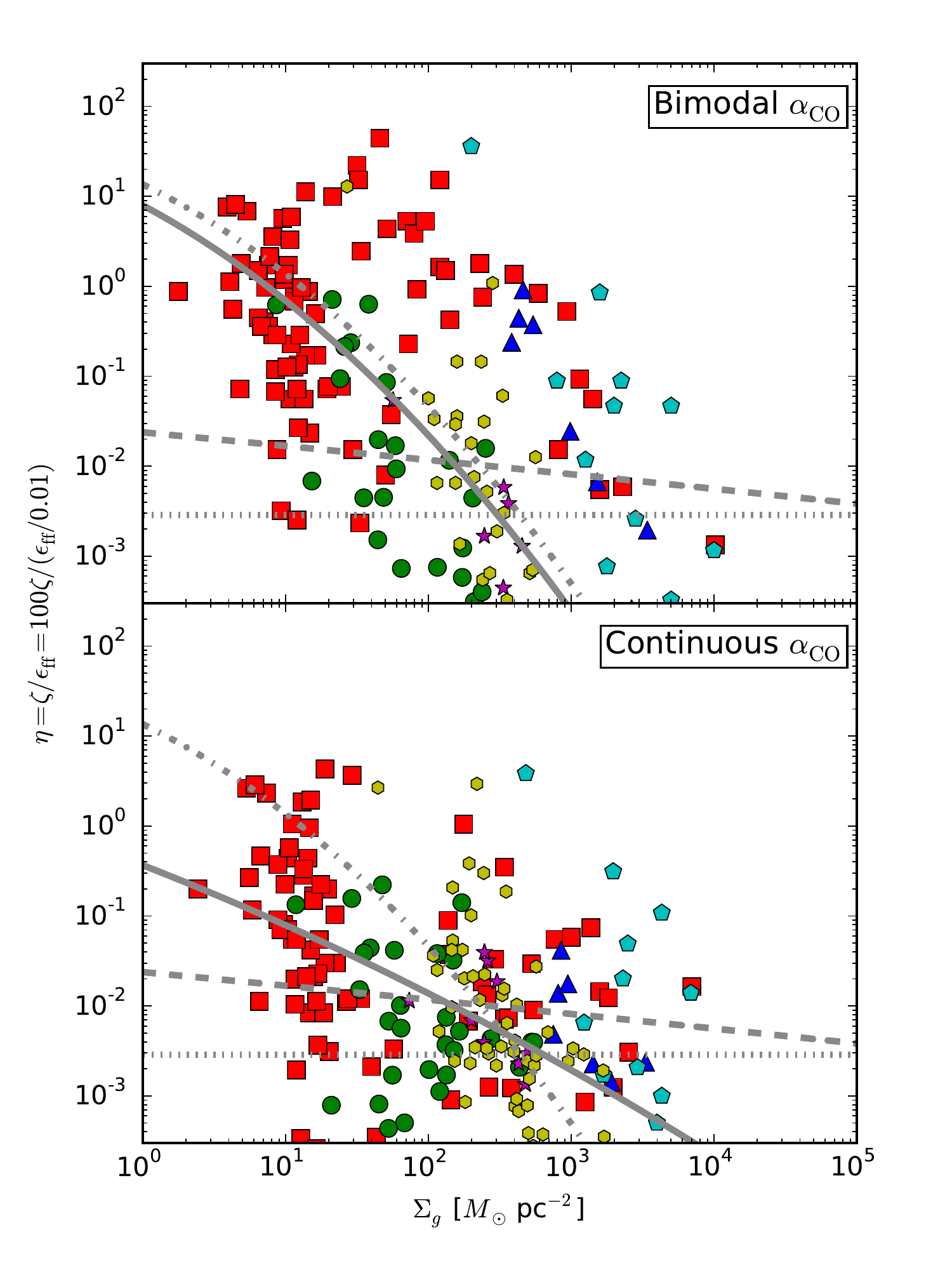}
}
\caption{
\label{fig:galaxies2}
Values of $\zeta$, the fraction of the mass that is super-Eddington, versus gas surface density $\Sigma_g$. Data points are identical to those shown in Figure \ref{fig:galaxies1}, and gray lines are for the same fits as in that Figure: the best-fit to the full data set (solid) the \citet{kennicutt98} relation (dot-dashed), and the fits by \citet{narayanan12a} (dashed) and \citet{faucher-giguere_feedback} (dot-dashed). As in Figure \ref{fig:galaxies1}, in the the upper panel the observed galaxies have surface densities assigned using the \citet{daddi10b} bimodal $\alpha_{\rm CO}$, while in the lower panel surface densities are computed using the theoretical $\alpha_{\rm CO}$ of \citet{narayanan12a}. All calculations are done for a Mach number $\mathcal{M} = 30$ and $\alpha=2.5$ for the slope of the density power spectrum in equation (\ref{eq:pofk}). Different choices change the results quantitatively but not qualitatively.
}
\end{figure}

In Figure \ref{fig:galaxies2}, we show the implications of these results for $\zeta$, the fraction of mass that is expected to be super-Eddington, and the mass-loading rate.  We parameterize the star formation rate per unit area as
\beq
\dot{\Sigma}_\star=\epsilon_{\rm ff}\Sigma_g/t_{\rm ff},
\eeq
where $t_{\rm ff}$ is the free-fall time at the average density of the star-forming gas and $\epsilon_{\rm ff}$ is an efficiency factor.  If we adopt as an \textit{ansatz} that the wind mass ejection rate is given roughly by
\beq
\dot{\Sigma}_{\rm wind}=\zeta(x_{\rm crit})\Sigma_g/t_{\rm ff},
\eeq
this implies that the mass loading factor
\beq
\eta\equiv\frac{\dot{\Sigma}_{\rm wind}}{\dot{\Sigma}_{\star}}=\frac{\zeta}{\epsilon_{\rm ff}}.
\label{eta}
\eeq
Figure  \ref{fig:galaxies2} shows $\eta$ for the galaxies in the sample of Figure \ref{fig:galaxies1} with a fiducial choice of $\epsilon_{\rm ff}=0.01$. Our values of $\zeta$, which range from $\sim 10^{-4}-10^{-1}$, coupled with our choice of  $\epsilon_{\rm ff}$ on galactic scales,  imply mass loading factors of $\sim 10^{-3}-10$.  Figure  \ref{fig:galaxies2}  suggests that the direct single-scattering radiation pressure may be a major contributor to the mass loading on galactic scales for some galaxies with mean surface densities $\lesssim100$ $M_\odot$ pc$^{-2}$. For galaxies with higher surface densities the effects of radiation pressure are systematically smaller because the surface density does not rise with star formation rate quickly enough to keep $F/F_{\rm Edd}$ from falling \citep{andrews_thompson}. Since $\zeta$ is exponentially sensitive to $F/F_{\rm Edd}$, even a modest fall translates to a dramatic reduction in the mass-loading factor.\footnote{At high surface densities, the effects of large FIR optical depth and the reprocessed radiation may increase the importance of radiation pressure in dense starbursts or their FIR optically-thick star-forming sub-regions, but we do not focus on this regime. See \cite{tqm,andrews_thompson,mqt10,krumholz_thompson12,krumholz_thompson13,davis_kt,skinner_ostriker}.}

Note that the choice of $\epsilon_{\rm ff}$ directly impacts the inferred mass-loading factor $\eta$, and for this reason we explicitly label the ordinate of Figure \ref{fig:galaxies2} with the $\epsilon_{\rm ff}$ dependence. Since we assume radiation pressure acts on local scales in individual star forming regions, we adopt a constant value for $\epsilon_{\rm ff}$ inferred from the free-fall time for the average of the star-forming molecular clouds on large scales \citep{krumholz14_review}. We extend this to feedback within GMCs with varying $\epsilon_{\rm ff}$ in Section \ref{section:gmcs}.  Note that our working definition of $\epsilon_{\rm ff}$ in Figure \ref{fig:galaxies2} is different from the one used by \cite{faucher-giguere_feedback} who evaluate the free-fall time at the average ISM density (typically significantly lower than the average density of star-forming clouds) and argue both that some galaxies have $\epsilon_{\rm ff} \gtrsim 0.1$, and that rather than being constant, $\epsilon_{\rm ff}$ scales with gas fraction and circular velocity. 

We caution that the results presented are also sensitive to the choice of the factor $\alpha_{\rm CO}$ used to convert between observable CO emission and surface density of molecular gas, particularly because of the exponential dependence of $\zeta$ on $F/F_{\rm Edd}$. To illustrate this, the top panels of Figures \ref{fig:galaxies1} and \ref{fig:galaxies2} show the data where we have inferred gas surface densities using the values of $\alpha_{\rm CO}$ suggested by \citet{daddi10b}: $\alpha_{\rm CO} = 4.6$ $M_\odot$ $(\mbox{K km s}^{-1}\mbox{ pc}^{-2})^{-1}$ for local non-starburst galaxies, $\alpha_{\rm CO} = 3.6$ $M_\odot$ $(\mbox{K km s}^{-1}\mbox{ pc}^{-2})^{-1}$ for high-$z$ disk galaxies, and $\alpha_{\rm CO} = 0.8$ $M_\odot$ $(\mbox{K km s}^{-1}\mbox{ pc}^{-2})^{-1}$ for starburst galaxies. The bottom panels show the exact same data but using the theoretical calibration of $\alpha_{\rm CO}$ computed based on simulations and radiative transfer post-processing by \citet{narayanan11a, narayanan12a}: $\alpha_{\rm CO} = \min[6.3, 10.7 \langle W_{\rm CO}\rangle^{-0.32}]/Z'^{0.65}$ $(\mbox{K km s}^{-1}\mbox{ pc}^{-2})^{-1}$, where $\langle W_{\rm CO}\rangle$ is the observed CO line intensity in units of K km s$^{-1}$, $Z'$ is the metallicity normalized to the Milky Way value, and we have used $Z'=1$ for all galaxies.

The effect of using the theoretical $\alpha_{\rm CO}$ calibration on the location of galaxies in the $\Sigma_g - \dot{\Sigma}_*$ plane is relatively modest, but it does lead to a somewhat steeper rise in $\dot{\Sigma}_*$ with $\Sigma_g$. Formally, the a linear least squares fit for the relation between $\Sigma_g$ and $\dot{\Sigma}_*$ gives
\beq
\log\dot{\Sigma}_* = 1.43\log\Sigma_g - 3.77
\label{daddi}
\eeq
for the \citet{daddi10b} $\alpha_{\rm CO}$, and
\beq
\log\dot{\Sigma}_* = 1.74\log\Sigma_g - 4.46
\label{narayanan}
\eeq
for \citet{narayanan11a, narayanan12a}; in these formulae the gas surface densities are in units of $M_\odot$ pc$^{-2}$ and the star formation rates are in units of $M_\odot$ pc$^{-2}$ Myr$^{-1}$, and the fits have been performed weighting all galaxies equally. Note that the slope of $1.74$ we find using the \citeauthor{narayanan12a}~calibration is shallower than the value of 1.95 found in the original \citeauthor{narayanan12a}~paper. This is due to the significantly-expanded data set we make use of here.  Using a variable $\alpha_{\rm CO}$, \cite{ostriker_shetty} also found a relatively steep correlation: $\log \dot{\Sigma}_*=1.9 \log\Sigma_g - 5.05$.

Although the change in slope between the two calibrations in equations (\ref{daddi}) and (\ref{narayanan}) is only about $0.3$, this has the effect of making $\zeta$ fall off much less dramatically with increasing $\Sigma_g$ using the \citet{narayanan11a, narayanan12a} calibration rather than that from  \citet{daddi10b}. The difference is not enough to render radiation pressure significant in starbursts, but it is a reminder that a slope of 2 in the $\log \dot{\Sigma}_*-\log \Sigma_g$ relationship represents a critical value for momentum feedback models \citep{tqm,andrews_thompson}. Changes in the $\alpha_{\rm CO}$ calibration severe enough to produce a star formation law with a slope of 2 appear unlikely based on current observations or theoretical models, but given the significant uncertainties can by no means be ruled out.  

\begin{figure}
\centerline{
\includegraphics[width=0.55\textwidth]{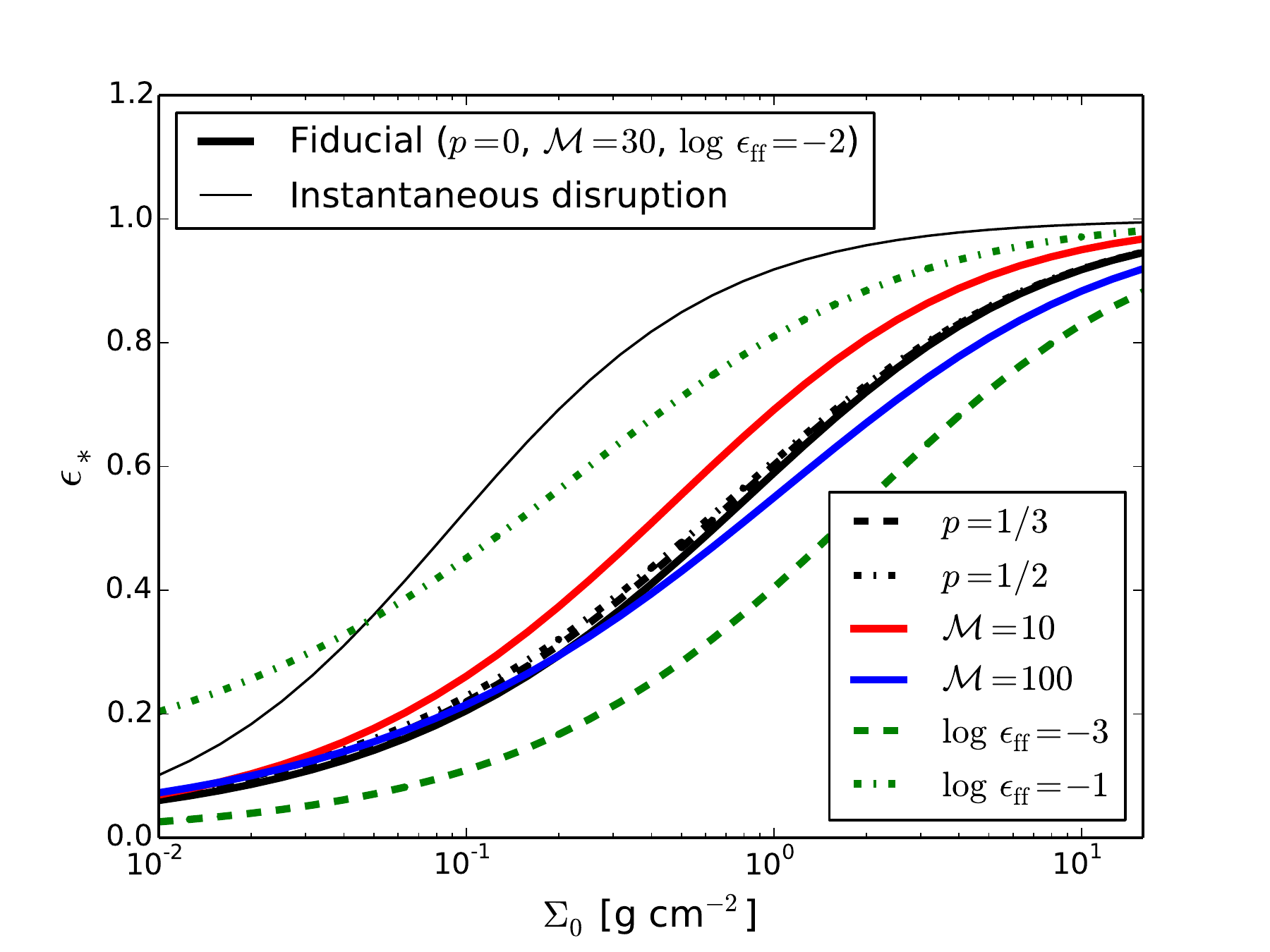}
}
\caption{Final star formation efficiency $\epsilon_*$ versus starting surface density $\Sigma_0$ for our simple model of radiatively-driven mass loss. The black solid curve shows a fiducial set of parameters $p=0$, $\mathcal{M}=30$, and $\epsilon_{\rm ff} = 0.01$, while the other curves show the results of varying one of these parameters.}
\label{figure:sfegmc}
\end{figure}

\subsection{Application to Giant Molecular Clouds}
\label{section:gmcs}

While the effects of radiation pressure alone may be fairly modest at galactic scales, they are much more significant at the scales of giant molecular clouds, which have much shallower potential wells than galaxies and proportionately higher star formation rates per unit mass. Consider a simple phenomenological model of a forming star cluster, somewhat similar to the models previously considered by \citet{mqt10}, \citet{fall10}, and \citet{dekel13}: we start with a spherical ball of gas with an initial mass $M_g(0)$ and radius $R(0)$, containing no stars. At time $t=0$ star formation begins, and thereafter occurs at a rate
\beq
\label{eq:sfr}
\dot{M}_* = \epsilon_{\rm ff} \frac{M_g}{t_{\rm ff}},
\eeq
where $M_g$ and $t_{\rm ff}$ are the instantaneous values of the gas mass and free-fall time.

The instantaneous stellar mass is $M_*$, and this stellar population produced a luminosity $L_* = \Psi M_*$, where $\Psi = 1140 L_\odot/M_\odot$ is the light to mass ratio of a zero-age stellar population with a standard IMF. Making the same \textit{ansatz} as in the previous section, we estimate that the light from the newborn stars drives a wind out of the system with a mass flux
\beq
\label{eq:windrate}
\dot{M}_w = \zeta(x_{\rm crit}) M_g/t_{\rm ff},
\eeq
where in this spherical geometry
\beq
\label{eq:xcritgmc}
\ln \,x_{\rm crit} = \left(\frac{1-f_g}{f_g}\right) \frac{\Psi}{4\pi G c \Sigma_{\rm tot}}.
\eeq
Here $f_g = M_g / (M_g + M_*)$ is the instantaneous gas fraction, and $\Sigma_{\rm tot}$ is the total gas plus stellar surface density. This last term will depend upon how the radius changes as star formation proceeds, and choose to parameterize this with a power law relationship, $R\propto [(M_*+M_g) / M_g(0)]^p$, where $p=0$ corresponds to the gas-plus stellar cloud maintaining a constant radius, $p=1/2$ to constant surface density, and $p=1/3$ to constant volume density. Consequently, the surface density $\Sigma_{\rm tot}$ evolves as
\beq
\label{eq:sigmatot}
\Sigma_{\rm tot} = \Sigma_0 \left(\frac{M_* + M_g}{M_g(0)}\right)^{1-2p},
\eeq
where $\Sigma_0 = M_g(0)/\pi R(0)^2$.

Dividing equations (\ref{eq:sfr}) and (\ref{eq:windrate}), we have
\beq
\frac{dM_w}{dM_*} = \eta = \frac{\zeta(x_{\rm crit})}{\epsilon_{\rm ff}},
\eeq
which we can easily integrate to obtain $M_w$ as a function of $M_*$, where $M_w$ is the mass ejected by winds up to the point where a mass $M_*$ of stars has formed. The final star formation efficiency is simply
\beq
\epsilon_* = \frac{M_*}{M_g(0)} = \frac{M_*}{M_* + M_w}
\eeq
evaluated at the point where $M_* + M_w = M_g(0)$, i.e., at the point where all the gas has been either converted to stars or lost to winds; $\epsilon_*$ is a function of the starting surface density $\Sigma_0$, the Mach number $\mathcal{M}$ of the turbulence, the index $p$ describing how the radius changes as star formation proceeds, and the star formation rate per free-fall time $\epsilon_{\rm ff}$. However, only the first of these matters to any substantial degree. Figure \ref{figure:sfegmc} shows $\epsilon_*$ versus $\Sigma_0$ for a range of choices for other parameters. As the plot shows, a generic result is that star formation efficiencies of $\sim 50\%$ are reached at surface densities of $\Sigma_0 \approx 1$ g cm$^{-2}$; this is consistent with the findings of \citet{fall10}, who used a much simpler model of mass loss that considered only explosive ejection of material and not steady winds as we have here.  The thin solid line labelled ``instantaneous disruption" assumes that the medium must come to the average Eddington limit before ejecting any mass and corresponds closely to the work of \cite{mqt10}.  At typical surface densities of galactic giant molecular clouds, $\sim 0.03$ g cm$^{-2}$ \citep{dobbs14}, the expected star formation efficiency is $\sim 10\%$, consistent with the low values typically found for such systems.

\subsection{Other Momentum Injection Sources}
\label{section:momentum}

In the application to whole galaxies, it is worth highlighting the potential importance of momentum injection by supernova explosions.  \cite{tqm} investigated the importance of momentum injection by supernovae in the ISM, and found them to be comparable to radiation pressure at high gas densities based on the work of \cite{thornton}.  Recent reappraisals by \cite{ostriker_shetty} and \cite{faucher-giguere_feedback} show that the net momentum input from supernovae can be as large as $\dot{P}_{\rm SNe}\sim10-20\times L/c$, with a weaker gas density dependence (see also \citealt{martizzi}).  However, the calculation of the net momentum input to a turbulent medium from supernovae is likely less straightforward in an analytic approach than radiation pressure because the momentum of supernova explosions --- the $10-20$ enhancement relative to $L/c$ --- is accumulated during the energy-conserving phase, as the remnants sweeps up mass.  Individual parcels of gas with column density $\Sigma$ (as in Section \ref{section:eddington}) will see a highly intermittent momentum injection rate.  More work is needed to understand how the momentum injection rate from supernovae couples to the turbulent ISM (see, e.g., \citealt{kim_ostriker}).  However, it is clear from the normalization of $\dot{P}_{\rm SNe}$ that it may dominate momentum input in galaxies and drive strong outflows: increasing the nominal momentum injection rate into galaxies by a factor of $10-20$ would lower all of the black solid lines by the same factor in the left panels of Figure \ref{fig:galaxies2}, making all galaxies near-Eddington and dramatically increasing their mass loading rates (right panels). 

The ram pressure of a very hot thermal gas component, as in the wind model of \cite{CC85}, may also deposit momentum in the medium.  If the energy injection rate within a radius $R$ is parameterized as $\dot{E}_{\rm hot}=\alpha\dot{E}_{\rm SN}$, where $\dot{E}_{\rm SN}$ is the energy injection rate of supernovae ($\sim10^{51}$\,ergs$/100\,{\rm yr}$, per M$_\odot$/yr of star formation) and if the hot gas mass outflow rate is $\dot{M}_{\rm hot}=\beta\,{\rm SFR}$, then the momentum injection rate for the hot gas at the surface of the star-forming medium is $\dot{P}_{\rm hot}\simeq\dot{M}_{\rm hot}V_{\rm hot}$, where $V_{\rm hot}(R)\simeq(2\dot{E}_{\rm hot}/\dot{M}_{\rm hot})^{1/2}$.  Comparing this to the momentum injection rate from radiation pressure in the single-scattering limit yields $\dot{P}_{\rm hot}/(L/c)\simeq3 \,(\alpha\beta)^{1/2}\,(7\times10^{-4}/\epsilon)$.  For order unity $\alpha\beta$, as is inferred for the hot gas in the wind of M82 by \cite{strickland_heckman}, this momentum source may dominate radiation pressure, and, like $\dot{P}_{\rm SNe}$, the contribution from $\dot{P}_{\rm hot}$ could shift the Eddington limit downward in Figure \ref{fig:galaxies2}.  Limits from X-ray observations indicate that $\beta\lesssim1$ for ${\rm SFR}\gtrsim10$\,M$_\odot$ yr$^{-1}$ galaxies \citep{zhang14}, and numerical calculations indicate that both $\alpha$ and $\beta$ may be functions of the gas surface density of galaxies \citep{creasey}.

Other momentum injection sources include stellar winds and cosmic rays.  The former is expected to be significantly less important than supernovae in a time-averaged stellar population in galaxies \citep{leitherer}, but may well be important in the early-time disruption of GMCs before any supernovae have occurred.  Estimates for the momentum input from cosmic rays indicate that they may also dominate radiation pressure in normal galaxies and possibly starbursts, depending on the CR scattering mean free path, the cosmic ray pion production rate, and --- of specific relevance for this work --- how the CRs interact and dynamically couple to low-column density regions in a turbulent ISM \citep{ptuskin,socrates_cr,jubelgas,lacki_gev,hanasz}.

\section{Conclusions}
\label{section:conclusion}

The Eddington luminosity per unit mass for gas subject to a momentum source is proportional to the column density of the medium.  Because turbulent media present a broad column density distribution to the momentum sources, there a exists a critical column density below which the medium is super-Eddington.  Even systems that are sub-Eddington will have super-Eddington sightlines.  

We have developed this idea using a simple formalism and discussed some of the implications for observed galaxies (Fig.~\ref{fig:galaxies2}) and the star formation efficiency, evolution, and disruption of GMCs (Fig.~\ref{figure:sfegmc}).  For average Eddington ratios of 0.1 and Mach numbers greater than about 30, we expect $\sim1$\,\% of the mass of the medium to be ejected per dynamical timescale (Fig.~\ref{figure:pdf}) with a well-defined velocity distribution (Fig.~\ref{figure:v}).   This rate of mass ejection is comparable to the star formation efficiency per free-fall time when averaged on large scales and thus the instability we identify here may be relevant for ejecting gas even when the system is significantly below the average Eddington limit.  An important uncertainty in our model is whether or not the projected persistence time of low column density structures in the tail of the column density PDF is shorter than the dynamical timescale.  This comparison of timescales directly affects whether material is ejected (Section \ref{section:turbulence}), and should be investigated in multi-dimensional simulations.

We discuss a number of momentum sources in Section \ref{section:momentum}, but focus our discussion on radiation pressure in the single-scattering limit, which is likely most relevant for the early-time disruption of GMCs.  Figure \ref{fig:galaxies1} shows the Eddington flux from normal galaxies and Figure \ref{fig:galaxies2} shows an estimate of their mass-loading factors $\eta$ (eq.~\ref{eta}). For galaxies with gas surface densities less than $\sim100$\,M$_\odot$ pc$^{-2}$, we find a large scatter in $\eta$ from $\sim10^{-3}-10$, indicating that radiation pressure feedback alone may be important for some galaxies. However, our results are highly sensitive to the assumed gas fraction and $\alpha_{\rm CO}$. On small scales, we compute the star formation efficiency in GMCs of specified initial gas surface density (Fig.~\ref{figure:sfegmc}).

The model is readily incorporated into subgrid and semi-analytic models of galaxies and/or GMC disruption.  The addition of other momentum sources like supernovae, hot winds, and cosmic rays in such models remains the subject of future work, but the estimates of Section \ref{section:momentum} suggest that they may be significant in driving material in low column-density super-Eddington sightlines out of star-forming galaxies and individual star-forming regions. 

\section*{Acknowledgments}
We thank Eve Ostriker and Norm Murray for comments and Brant Robertson, Evan Scannapieco, Eliot Quataert, and Phil Hopkins for useful discussions.  TAT thanks Chris Kochanek for a reading of the text.  We gratefully acknowledge the Simons Foundation for funding the workshop {\it Galactic Winds: Beyond Phenomenology}, where this work was conceived.   We also thank the Kavli Institute for Theoretical Physics and the organizers of {\it Gravity's Loyal Opposition: The Physics of Star Formation Feedback}, where a portion of this paper was written.   This research was  supported in part by the National Science Foundation under Grant No.\ PHY11-25915.  TAT is supported in part by NASA Grant NNX10AD01G and NSF Grant 1516967.   MRK is supported by NSF grants AST-0955300 and AST-1405962, NASA ATP grant NNX13AB84G, NASA TCAN grant NNX14AB52G. 

\end{document}